# PARTICLE ACCELERATION IN SOLAR FLARES WITH IMAGING-SPECTROSCOPY IN SOFT X-RAYS

September 6, 2022

Mitsuo Oka, University of California Berkeley, moka@berkeley.edu
Amir Caspi, Southwest Research Institute
Bin Chen, New Jersey Institute of Technology
Mark Cheung, Lockheed Martin Solar and Astrophysics Laboratory
James Drake, University of Maryland
Dale Gary, New Jersey Institute of Technology
Lindsay Glesener, University of Minnesota
Fan Guo, Los Alamos National Laboratory
Hantao Ji, Princeton University
Xiaocan Li, Dartmouth College
Takuma Nakamura, University of Graz
Noriyuki Narukage, National Astronomical Observatory of Japan
Katharine Reeves, Harvard-Smithsonian Center for Astrophysics
Pascal Saint-Hilaire, University of California Berkeley
Taro Sakao, Japan Aerospace Exploration Agency
Chengcai Shen, Harvard-Smithsonian Center for Astrophysics
Amy Winebarger, NASA Marshall Space Flight Center
Tom Woods, University of Colorado

## ABSTRACT

Particles are accelerated to very high, non-thermal energies during explosive energy-release phenomena in space, solar, and astrophysical plasma environments. In the case of solar flares, it has been established that magnetic reconnection plays an important role for releasing the magnetic energy, but it remains unclear if or how magnetic reconnection can further explain particle acceleration during flares. Here we argue that the key issue is the lack of understanding of the precise context of particle acceleration but it can be overcome, in the near future, by performing imaging-spectroscopy in soft X-rays (SXRs). Such observations should be complemented by observations in other wavelengths such as extreme-ultraviolets (EUVs), microwaves, hard X-rays (HXRs), and gamma-rays. Also, numerical simulations will be crucial for further narrowing down the particle acceleration mechanism in the context revealed by the observations. Of all these efforts, imaging-spectroscopy in SXRs, if successfully applied to large limb flares, will be a milestone in our challenge of understanding electron acceleration in solar flares and beyond, i.e. the Plasma Universe.

# 1. THE BIG PICTURE

Since the discovery of galactic cosmic rays in 1912 by a Nobel laureate Victor F. Hess, the physics community has been trying to understand why and how charged particles are accelerated to very high, non-thermal energies in outer space. In 1986, another Nobel laureate, Hannes Alfvén coined the term *Plasma Universe* as he envisioned that knowledge expands from plasma experiments in the laboratory to the magnetospheres and also astrophysics in general (Alfvén, 1986). After decades of study, we now know plasma particles are indeed accelerated in various space, solar, and astrophysical plasma environments. However, it remains unclear how particles are accelerated in each environment and if (or what) universal scaling laws exist. This applies particularly to particle acceleration in explosive energy-release phenomena such as solar/stellar/pulsar/magnetar flares as well as terrestrial/planetary substorms. While a standard theory exists for particle acceleration at shocks (such as interplanetary shocks and supernova remnant shocks), a plethora of theories have been proposed to explain particle acceleration in explosive energy-release, and they have not been constrained.

While this problem of **Particle Acceleration in the Plasma Universe** has remained unsolved for many decades, it has become more clear recently that the solar plasma environment is in a strategically important location in the plasma-parameter space (Ji & Daughton, 2011) (See also Fig. 1). In the past decades, the fluid approach (e.g. MHD simulations) has been advancing by increasing the Lundquist number $S$ (green arrow). The kinetic approach (e.g. particle-in-cell simulations) has been advancing by increasing the system size $\lambda$ normalized by the kinetic scale (orange arrow). These approaches are very roughly aligned with a shift from the space plasma regime to the solar coronal plasma regime (orange arrow) as well as the historical scale-up of the laboratory experiments (pink arrow). If we can have a breakthrough in our understanding of the plasma processes in the solar corona (gray filled circle) with all these multiple lines of approach, it would open up the avenue toward the vast region of parameter-space where various astrophysical plasma environments reside (cyan arrows). Therefore, the solar coronal plasma environment has received considerable attention from the laboratory, space, and astrophysical plasma communities, although many comparative studies have been already carried out from earlier years (e.g., Kennel et al. 1985; Terasawa et al. 2000; Bhattacharjee, 2004; Reeves et al. 2008; Zweibel & Yamada, 2009; Oka et al. 2018; See also white paper by Vievering et al. 2022). In fact, it has been recognized that solar *remote-sensing* observation is special in a sense that, unlike *in-situ* measurements of Earth's magnetosphere, it provides large-scale contexts through imaging, and yet it still resolves spatial and temporal structures of the phenomena in much more detail than unresolvable observations of distant astrophysical objects.

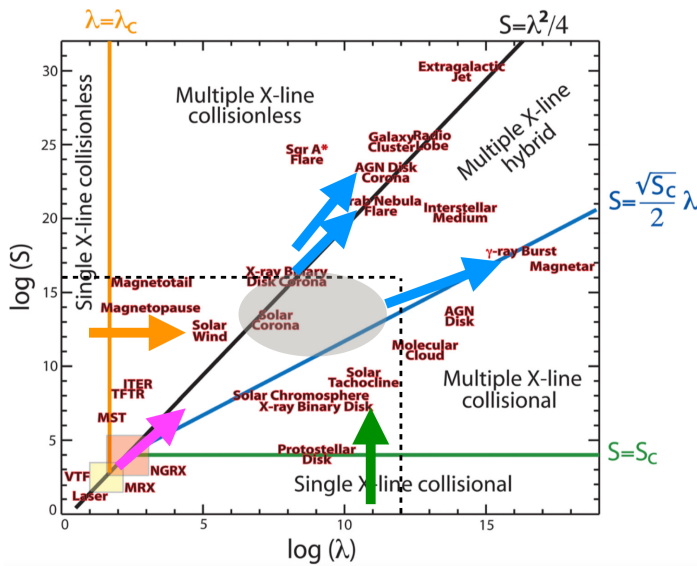

**Figure 1 | 2D parameter dependence of magnetic reconnection and various plasma environments.** While reconnection can occur in laboratory, space, solar, and astrophysical plasma environments, they exhibit different properties depending on the Lundquist number S and the system size $\lambda$ in units of ion inertial length. In the solar corona, magnetic reconnection is likely dominated by multiple X-line with a mixture of collisional and collisionless dynamics. Reproduced from Ji and Daughton (2011), with the permission of AIP Publishing.

## 2. THE CHALLENGE

Despite the advantages of solar remote-sensing observations (especially imaging), the problem of particle acceleration in solar flares remains unsolved. It has been established that magnetic reconnection plays an important role for releasing the magnetic energy, but it remains unclear if or how magnetic reconnection can further explain particle acceleration during solar flares. As will be described below, we consider that **the key issue is the lack of understanding of the precise context of particle acceleration**.

Figure 2 illustrates the challenge. In the EUV range (typically < 0.5 keV), the large-scale loops and associated magnetic field configurations can be easily captured (Fig. 2a). More detailed analyses, often combined with DEM analysis and spectroscopy, have further indicated a variety of reconnection-related plasma structures such as plasmoids, jets, converging inflows, turbulence, termination shock, etc. These structures have been successfully reproduced by magnetohyro-dynamic (MHD) simulations and have served to establish the reconnection paradigm of solar flare dynamics.

However, such structures are seen only in rare cases of EUV observations. To date, a typical EUV image would appear like the one in Fig. 2a which does not show the key structures such as shocks and plasmoids, making it difficult to interpret the signatures of accelerated electrons as diagnosed by X-rays and microwaves. A likely explanation for the poor visibility of the important plasma structures in EUV is that the process of EUV emission involves both ionization and recombination, and it takes time for the ions to reach an equilibrium before producing EUV emissions (e.g. Imada et al. 2011; Shen et al. 2013; See also Fig. 2d). Such a time scale is comparable to the typical time scale of previous hard X-ray measurements (4 – 200 seconds) and is likely to be much longer than the acceleration time scale. Thus, EUV diagnostics of the coronal plasma can be confounded by such an effect of delayed emission.

X-ray continuum, on the other hand, are produced via the bremsstrahlung emission without any delay. Previous observations showed that the non-thermal, hard X-ray (HXR) signals (typically >10 keV) come primarily from the chromosphere at the footpoints of the flaring loop (blue contours in Fig. 2c), while the lower-energy (typically < 10 keV), thermal component comes from the corona at and around the top of the flaring loop (red contours in Fig. 2c). The location of this thermal component (red contours in Fig. 2c) matches with the location of the looptop region seen in EUV (the dark green region in Fig. 2a). The higher-energy, non-thermal emissions can also come from the corona somewhat 'above' the mostly thermal looptop region (black contours, Fig. 2c), indicating the location of energy-release can be further above and away from the looptop region (e.g. Masuda et al. 1994; Krucker et al. 2010; Sui & Holman, 2003).

However, the X-ray intensity is very low in the 'above-the-looptop' (ALT) region and almost zero (or within the background level) in the presumed energy-release site (toward the right of the black contours in Fig. 2c; See also Fig. 3 for expected structures) mainly because of the rather low density in these regions. The intensity of the bremsstrahlung emission depends on the density of the local ion population. This low intensity of the coronal X-ray emission makes it difficult to diagnose electron acceleration with high temporal and spatial resolutions.

It is to be emphasized that theories expect various features in the solar corona such as shocks, turbulence, collapsing magnetic structures, plasmoids, and turbulence. These structures have been used in proposed theories of particle acceleration mechanisms (e.g. Aschwanden, 2005). For example, some theories use the termination shock (e.g. Tsuneta & Naito, 1998; Guo and Giacalone, 2012) while some other theories use magnetic reconnection and associated plasmoids to explain particle acceleration (e.g. Drake et al. 2006; Oka et al. 2010; Li et al. 2015; Arnold et al. 2021). Also, multiple structures are often combined to explain particle acceleration (e.g., Shen et al. 2018; Kong et al. 2019). These structures, especially shocks and magnetic reconnection, are fundamental processes in plasmas, but we still do not know which process is more important for particle acceleration in solar flares. Because of the large uncertainty in both EUV and hard X-ray observations and hence the lack of our understanding of the precise context of particle acceleration, the particle acceleration mechanism is not constrained.

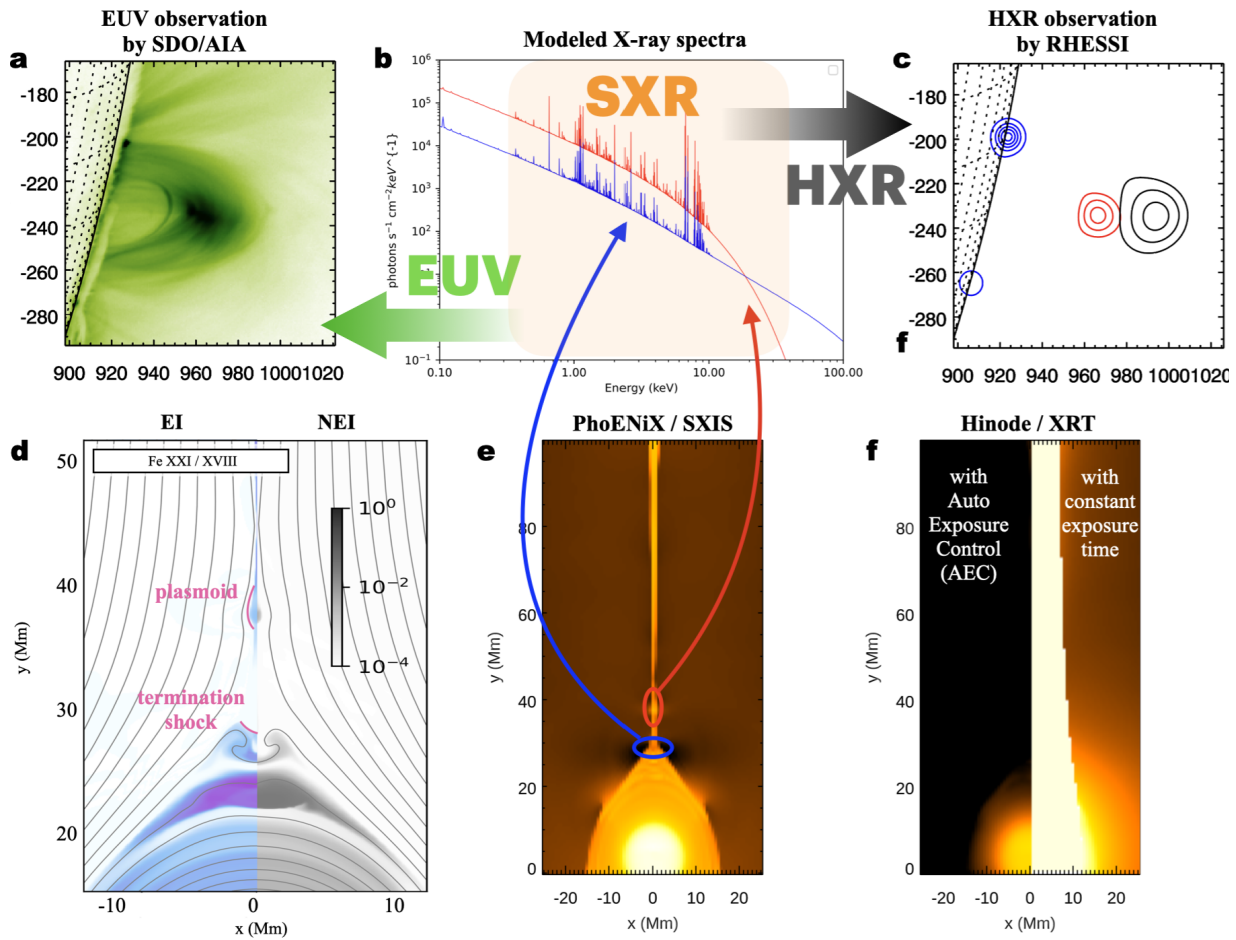

**Figure 2 | Example observations and simulations, demonstrating the advantages of using soft X-rays to diagnose solar flares.** (a,c) EUV and HXR images from a flare on July 19, 2012. Reproduced from Krucker and Battalia (2014) by Oka et al. (2015) © AAS. Reproduced with permission. (b) Modeled X-ray spectra from hypothetical electron distributions. (d) Line intensity ratio from MHD simulations with equilibrium ionization (EI) and non-equilibrium ionization (NE) (Shen et al. 2013). (e,f): Simulated soft X-ray (SXR) images of the flaring reconnection region by the PhoENiX mission concept and currently-operating Hinode/XRT.

In other words, there is a gap between the flare-scale dynamics as can be diagnosed mainly with EUV imaging and the signatures of accelerated particles as can be diagnosed mainly with hard X-ray imaging. This gap can be viewed as a gap in the energy coverage, as shown in Fig. 2b. In between these ranges is the soft X-ray (SXR) range that represents the heated, high-temperature plasma in the corona. The SXR range is crucial for our understanding of how the electron energy transitions from thermal to non-thermal. This is in contrast to the HXR range that provides the information of electrons that are already accelerated and transported to the dense location where sufficiently intense X-ray emissions can be produced.

## 3. PROPOSED SOLUTION

To overcome the challenge described above and to fill the gap between the EUV and HXR energy ranges, **we recommend studying large-scale, limb flares through imaging-spectroscopy in the soft X-ray (SXR) range (0.5 - 10 keV) with a sufficiently large dynamic range as well as high temporal, spatial, and energy resolutions.**

<u>Imaging-spectroscopy</u> Imaging-spectroscopy refers to a methodology in which an energy spectrum is obtained at each location in an image (as illustrated in Fig. 2b and 2e). If we capture different structures such as shocks, plasmoids, and collapsing regions all in the same field of view via SXR imaging, follow their time evolutions, and simultaneously obtain energy spectra at each structure, then we would be able to (1) quantify various parameters at each structure, (2) identify which structure is more important for electron acceleration, and (3) constrain the acceleration mechanism. Such observations with SXRs should be supplemented with hard X-ray (HXR) measurements that extend up to ~100 keV, in order to evaluate various parameters including, but not limited to, temperature, emission measure (density), power-law index, the non-thermal fraction of electron energies, and the low-energy and high-energy cutoffs.

<u>High sensitivity and large dynamic range</u> The imaging-spectroscopy in SXRs needs to achieve a higher sensitivity and larger dynamic range compared to existing SXR imaging. As described in the previous section, emissions from the ALT region and the energy-release site are very faint compared to the bright sources on the flaring loop, and the conventional observations have not been able to study such faint sources in detail due to the limited dynamic range. The situation is illustrated in Fig. 3f. While Hinode/XRT – a currently-operating, SXR bandpass imager *without* spectroscopy – can automatically adjust the exposure time, it sometimes captures the brightest region only because of the limited dynamic range. The exposure time can be manually extended to gain photon counts from the faint regions (such as the ALT and energy-release site), but then the photon counts would be saturated in bright regions. Such problems can be avoided by increasing the dynamic range as shown in Fig. 3e (More specifically, we would need $>10^4$ for 0.5 – 10 keV and $>10^3$ for 10 – 30 keV).

Spatially large, limb flares The imaging-spectroscopy observations should focus on (though are not limited to) solar flares that occur near the edge of the solar disk, the solar limb. Such *limb events* allow us to view solar flares from the side (as in Fig. 2) and therefore separate the different spatial structures and sources without foreshortening by projection. Also, because the basic flare configuration and associated structures should be similar for flares with different spatial sizes, we should focus on spatially large flares in order to maximize the output from a feasible spatial resolution and better identify the flare-associated structures such as the termination shock and plasmoids. Such spatially-large flares are often intense and fall into the GOES M-class or larger.

Spatial resolution A similar argument can be used to derive the requirement for the spatial resolution in future observations. For example, plasmoids can have different sizes. Some plasmoids may be too small to be captured by a telescope but some other plasmoids may be sufficiently large to be captured by the same telescope. However, above the collision scale (i.e., $\geq$ 0.1 arcsec ~ 0.1 Mm), the magnetohydro-dynamics (MHD) can be considered scale-free, and the basic laws of physics that govern the plasmoid dynamics should apply equally to the plasmoids of different sizes. Thus, while it is always preferable to have a higher spatial resolution, we can learn significantly from the observations of large plasmoids alone. Previous EUV observations have captured plasmoids with the spatial resolution of 1.5 arcsec (e.g. Takasao et al. 2012), and so future imaging-spectroscopy in SXRs should first aim at a similar spatial resolution to achieve the next breakthrough.

Energy resolution There are many spectral lines in the soft X-ray range (Figure 2b) and they provide valuable information on the thermal/ionization equilibrium and abundances of different particle species in addition to more typical parameters such as temperature and density. While individual lines should be ultimately resolved for full diagnostics, we require the energy resolution to be <0.2 keV FWHM for the 0.5 – 10 keV range for achieving the next breakthrough. This is sufficient to distinguish between prominent line clusters such as the (cool) Fe 6.4 keV and (hot) Fe 6.7 keV emission. The 6.4 keV line comes from neutral or weakly ionized Fe ions at the photosphere, but the emission is produced by inner-shell ionization via high-energy (> 6.4 keV) electrons and could be used to detect and diagnose high-energy, *non-thermal* electrons as the temperature of ambient electrons is substantially lower than 6.4 keV. Phillips et al. (2012) estimated the Fe abundance ratio in the photosphere from the line complex measured at ~6.4 keV.

# 4. PROPOSED IMPLEMENTATION

**The proposed solution, i.e. imaging-spectroscopy in soft X-rays (SXRs), can be achieved by a photon-counting technology combined with focusing optics**, although it should still be complemented by observations in other wavelengths such as extreme-ultraviolet (EUV), microwaves, hard X-rays (HXRs), and gamma-rays. Numerical simulations will also be crucial for narrowing down the particle acceleration mechanism in the context revealed by these observations.

<u>SXR photon-counting with focusing optics</u> The photon-counting technique with focusing optics will greatly improve sensitivity and dynamic range (e.g., Glesener et al. 2022). Also, with the photon-counting technique, the energy, location, and time of each and every detected photon are recorded so that scientists can choose any type of binning in any axis (time, space, and energy) before performing imaging-spectroscopy. The technology has been implemented in recent astrophysical missions such *NuSTAR* (for 2 - 80 keV) and is already being developed for solar-dedicated mission concepts such as *FOXSI-SMEX* (for 3 - 50 keV; Christe et al. 2022) and *FIERCE* (Shih et al. 2022). For the slightly lower, soft X-ray range of 0.5 - 10 keV, the number of photons will increase significantly, posing a challenge to the development of sensors and mirrors. We recommend NASA to invest in the development of such technologies and/or facilitate international collaborations. For example, *PhoENiX*, a SMEX-size mission concept proposed to JAXA in 2022, includes the solar-dedicated, SXR photon-counting with focusing optics (Figure 2b, 2e). Thus, JAXA could be a possible partner for collaboration. In fact, there have already been intense US-Japan collaborations in the NASA-funded *FOXSI* rocket series (Glesener et al. 2022). It is also important to invest in complementary approaches such as microcalorimeter and slitless imaging-spectrograph. These technologies can achieve a high energy-resolution. While microcalorimeter is technologically challenging, the slitless imaging-spectrograph (with limited imaging capability) is already developed for the sounding rocket project *Marshall Grazing Incidence X-ray Spectrometer* (*MaGIXS*, launched in 2021) and the *CubeSat Imaging X-ray Solar Spectrometer* (*CubIXSS*, selected by NASA in 2021).

<u>Coordinated observations with other wavelengths</u> While imaging-spectroscopy in SXRs would enable us to study the precise context of electron acceleration, it needs to be combined with observations in other wavelengths in order to advance our understanding of electron acceleration during flares. Of particular importance is imaging-spectroscopy in hard X-rays (HXRs), as is already described above and also proposed in the *FOXSI-SMEX* and *FIERCE* concepts. If combined with SXR measurements, it would allow us to study electron acceleration seamlessly from thermal to non-thermal ranges. Imaging in EUVs is also important as the spatial resolution can be higher than that of imaging in X-rays and would provide a broader context of electron acceleration. The recently selected EUV missions such as *MUSE* (Cheung et al. 2022) and *Solar-C* (EUVST) are the examples that can serve for such a purpose. On the higher energy end of the spectrum, observations in microwaves and γ-rays would enable us to diagnose energetic electrons from ~100 keV to MeVs. Ground-based microwave observations by, for example, the currently-operating *EOVSA* and the proposed *FASR* project (Gary et al. 2022), enable imaging-spectroscopy and further provide additional information such as the magnetic field magnitude and angle to the line of sight (see white paper by Chen et al. 2022). The magnetic field magnitude is crucial for deriving the plasma beta, while the longitudinal component (the guide field) governs the efficiency of particle acceleration (e.g., Arnold et al. 2021). Measurements in the γ-ray range are also important as they can provide polarization information which is useful for understanding the anisotropy of accelerated electrons and thus for constraining the acceleration mechanism.

<u>Interpretation with numerical simulations</u> Understanding the precise context of electron acceleration with imaging-spectroscopy in SXRs, combined with other wavelengths, is already a large step forward, but the particle acceleration occurs in the kinetic scale (~1 m) which is many orders of magnitude smaller than the global flare size (~10 Mm) or the size of macro-scale structures such as plasmoids (< 1 Mm or 1.5 arcsec). Thus, it would be extremely challenging to fully constrain the acceleration mechanism with future observations alone. As such, conventional simulations with magnetohydro-dynamic (MHD), particle-in-cell (PIC), and other techniques will continue to be important for interpreting observations from the fluid to kinetic scales. A multi-scale approach would also be helpful. For example, in the recently developed *kglobal* code, particles with a guiding-center approximation are tracked in a 2D macrosystem that includes the self-consistent feedback of energetic particles on the dynamics (e.g. Drake et al. 2019; Arnold et al. 2021). A caveat is that all these types of simulations require a large computational resource for realistic modeling (e.g., Cheung et al. 2019; Chen et al. 2020; Li et al. 2021, 2022), and such realistic modeling is crucial for interpreting observation data and thus for maximizing the scientific output from a new solar mission with imaging-spectroscopy in SXRs. Therefore, we recommend NASA to provide sufficient funds for preparing and updating modeling tools (see also a white paper by Guo et al. 2022) prior to the launch of a new mission that would carry out imaging-spectroscopy, which should be distinct from the Guest Investigator Program. Such a pre-launch modeling would also be helpful for improving the mission design.

## 5. EXPECTED SIGNIFICANCE

This paper is motivated by the long-standing problem of particle acceleration in solar flares, and the proposed solution of imaging-spectroscopy in SXRs, if successfully implemented, will have a significant impact on the various problems of solar physics. Examples include, but are not limited to, the problem of energy partition (budget) during flares, the number problem of solar flares, the fundamental physics of magnetic reconnection, coronal mass ejection and associated production of Solar Energetic Particles (SEPs), coronal heating, microflares, and space weather. Moreover, while SXRs are produced by electrons, the precise context to be obtained by SXR imaging-spectroscopy would be helpful for our better understanding of not only electron acceleration but also ion acceleration. Also, as described in Section 1, plasma parameters in the solar corona have strategic importance. Thus, we expect that our proposed solution of solar flare studies would give a significant and interdisciplinary impact to other communities such as laboratory, space, and astrophysics.